\documentstyle[12pt]{article}

\begin{document}

\tolerance=1600

\begin{titlepage}
\begin{flushright}
\hfill YCTP-P10-98\\
\hfill hep-th/9804081
\end{flushright}
\vskip 1.5 cm

\begin{center} 
{ \LARGE \bf  Nahm Equations and Boundary Conditions}
\end{center}

\vskip .1cm

\begin{center}
Dimitrios Tsimpis \footnote[1]{\mbox{e-mail: tsimpis@yale.edu}}
\end{center}

\vskip .7cm

\begin{center}
{\it Yale University} \\
{{\it Department of Physics, Sloan Physics Laboratory} \\
     {\it PO Box 208120, New Haven, CT 06520-8120, USA}}
\end{center}

\vskip .2in

\begin{center} {\bf Abstract } \end{center}

We derive certain boundary conditions in Nahm's equations by considering a 
system of $N$ parallel D1-branes perpendicular to a D3-brane in type IIB string 
theory.

\vskip 6 cm
\end{titlepage}
\vfill
\eject

\newpage

\clearpage

\section{Introduction}

Nahm's equations for $SU(N)$ monopoles [1] were derived using D-brane techniques 
(for a review see eg. [2]) for the first time in [3]. Nahm's data however are 
incomplete without specifying the boundary conditions (bc) [4] and a D-brane 
derivation of those was lacking. 
Steps in this direction have been made in [5,6]. 

In this paper we show how Nahm's bc arise naturally in the context of D-brane 
physics by considering (following [3]) a system in type IIB string theory of $N$ 
infinite parallel D1-branes perpendicular to a D3-brane. The present analysis is 
relevant only to the case of discontinuous but finite Nahm's data. 
	
In order to derive the advertised result we consider a certain reduction / 
truncation of the low-energy effective action on the world-volume of the 
D1-branes. Nahm's equations and bc then follow from the requirement of a 
supersymmetric vacuum. 

The discontinuities in Nahm's data are encoded in the vacuum expectation values 
of the hypermultiplets, coming from the 1-3 sector, which appear as source terms 
localized on the intersection of the D1/D3 branes. This is the main result of 
this paper and is contained in Eq. (8) below.  Similar results have recently 
been obtained by Kapustin and Sethi [7].

\section{Analysis}

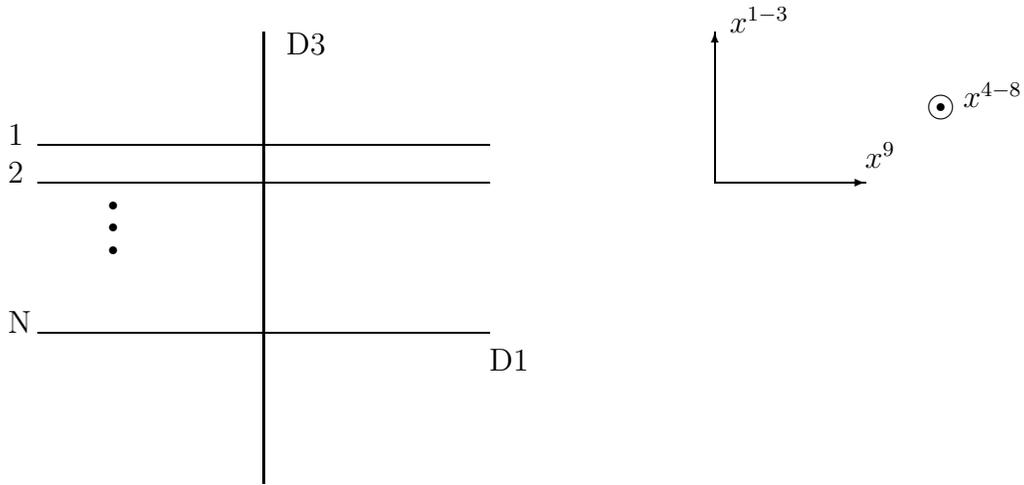
Consider the configuration, shown in fig.1, of a D3-brane extended in 
$x^{0,1,2,3}$ intersecting a system of $N$ infinite parallel D1-branes extended 
in 
$x^{0,9}$, at the point $x^{9}=0$ of the D1-branes' world volume. Let $Q_{L,R}$ 
be the supercharges associated to left, right-moving degrees of freedom of the 
IIB world-sheet theory. A D3-brane is invariant under supersymmetry 
transformations $\epsilon_{L}Q_{L}+\epsilon_{R}Q_{R} $ such that
\begin{equation}
\epsilon_{L}=\Gamma^{0123}\epsilon_{R}~~,
\end{equation}
where $\epsilon_{L,R} \sim 16_{+}$ of Spin(1,9). Similarly 
the D1-brane imposes the condition 
\begin{equation} 
\epsilon_{L}=\Gamma^{09}\epsilon_{R}~~. 
\end{equation}
The above two conditions imply
\begin{equation}
\epsilon_{L}=\Gamma^{1239}\epsilon_{L}~~.
\end{equation}
Hence the supersymmetry parameter $\epsilon_{L}^{\alpha A}$ must be in the 
($2_{+},4_{+}$) of $Spin(4)_{1239}\otimes Spin(1,5)_{04-8}$  where $A (\alpha)$ 
is a $Spin(4)  (Spin(1,5))$ index. In addition $\epsilon_{L}^{\alpha A}$ must 
satisfy 
the ``$SU(2)$ Majorana condition'' 
\begin{equation}
\varepsilon^{AB}C^{\alpha}_{ \beta}\epsilon_{L}^{\beta 
B}=(\epsilon_{L}^{\alpha A})^{\ast}~~,
\end{equation}
where $\varepsilon^{AB}$ is the rank-2 antisymmetric tensor and $C$ is the 
charge 
conjugation matrix in ${\bf R}^{1,5}_{04-8}$ (this comes from the Majorana condition in 10d). Therefore the configuration of fig.1 leaves 8 real 
supercharges 
unbroken. Note that the original ten-dimensional $Spin(1,9)$ invariance is 
broken 
down to $Spin(3)_{123} \otimes Spin(5)_{4-8}$. 

It will be convenient to parametrize the unbroken supersymmetry by a pair of 
chiral $Spin(4)_{1239}$ spinors $\eta_{i}$ $i=1,2$, transforming as a doublet of 
$SU(2)$ R-symmetry. $SU(2)_{R}$ can be thought of as $Spin(3)_{678}$ (see 
below). 

The low-energy fields on the D1-branes' world-volume can be found by quantizing 
the different string-theory sectors [2,8]: The 1-1 strings give bosonic fields 
$X^{M}_{mn}$, $M=0,...9$, transforming as a vector of $SO(1,9)$ and a 
Majorana-Weyl fermion $\psi_{mn}$ in 10d. These fields are in the adjoint of $SU(N)$: 
$m,n=1,...N$. The 1-3 sector gives a spinor of $Spin(4)_{1239}$ and a spinor of 
$Spin(1,5)_{04-8}$, both in the fundamental of $SU(N)$. (A GSO projection matching bosonic and fermionic degrees of freedom should be imposed). There are also fields coming from the 3-3 
sector but these are external and will not be taken into account here. 

We will consider the situation where the branes do not fluctuate along 
$x^{6,7,8}$ 
and we will set $X^{6,7,8}=0$. We also gauge away the longitudinal component 
$X^{0}$ 
of the gauge field of the D1-branes' world-volume. Moreover, anticipating Nahm's 
equations, we will require that the fields be time-independent and that they 
depend only on $x^{9}$, which we denote by $s$.  The original 10-dimensional 
spacetime symmetry is thus further broken down  to a {\it global}  
$Spin(3)_{123} \otimes Spin(2)_{45} \otimes Spin(3)_{678}$.

The fields $X^{4}, X^{5},  X^{\mu}$,   $\mu=1,2,3,9$, constitute the bosonic 
part 
of  the Yang-Mills multiplet of  the Euclideanized version of ``$N=2$ matter'' 
theory 
in 4d [9] reduced to one spatial dimension, namely $x^{9}$. Of course upon 
dimensional reduction only the $Spin(3)_{123}$ subgroup of the $Spin(4)_{1239}$ 
symmetry of the 4d 
theory survives as a (global) symmetry of the low-energy effective action. The 
full field 
content is organized as follows:

 {\bf Yang-Mills multiplet}: $X^{9}$ the world-volume ``gauge field''; 
$X^{a}_{mn}$  
$a=1,2,3$, a vector of $SO(3)_{123}$; $X^{4}_{mn}, X^{5}_{mn} $ an $SO(2)_{45}$ 
doublet of real bosons; $\lambda_{imn}$  $i=1,2$, an $SU(2)_{R}$ doublet of  
$Spin(3)_{123}$ spinors. All the fields in the Yang-Mills multiplet are in the 
adjoint 
of $SU(N)$:  $m,n=1,...N$.
 
{\bf Hypermultiplets}: An $SU(2)_{R}$ doublet of complex bosons $h_{im}$  $i=1,2 
$; a spinor $\chi_{m}$ of $Spin(4)_{1239}$ (which should really be thought of as 
a 
pair of $Spin(3)_{123}$ spinors). All the matter fields are in the fundamental 
of 
$SU(N)$: $m=1,...N$.

The hypermultiplets live on the intersection of the D1/D3-branes and are thus 
localized at the point s=0 on the D1-branes' world-volume.

For an off-shell realization we have to include the auxiliary bosonic fields 
$D_{ijmn}=D_{jimn}=(D_{i'j'mn})^{\ast}\varepsilon_{ii'}\varepsilon_{jj'}$ (a 
triplet of $SU(2)_{R}$ in the adjoint of $SU(N)$) and $F_{im}$ (an $SU(2)_{R}$ 
doublet in the fundamental of $SU(N)$). 

The low-energy effective Langrangian reads (after setting $g_{YM}=1$):
\begin{equation}
L=L_{kinetic}(X^{9},X^{a},X^{4},X^{5},\lambda_{i})+L_{0}
+\delta(s)\{L_{kinetic}(h_{im},\chi_{m})+L_{interaction}\}
\end{equation}
where 
\begin{equation}
L_{0}=\frac{1}{2}Tr\{(D^{2}-([X^{4},X^{5}])^2\}+fermions
\end{equation}
and
\begin{equation}
L_{interaction}=\mid F_{im}\mid 
^{2}+h_{im}^{\ast}[(X^{4})^{2}+(X^{5})^{2}]_{mn}h_{in}+D_{ijmn}h_{im}^{\ast}h_{j
n}
+fermions
\end{equation}

The Langrangian is invariant under SUSY transformations parametrized by 
$\eta_{i}$ \footnote[1]{in fact the $\delta(0)^{2}.0$ term in 
the on-shell SUSY variation of 
$L$ is ambiguous. The $\delta$ function should be thought of as receiving some 
kind of stringy regularization. We thank Greg Moore for pointing this out and 
for discussions on this issue.}. For a supersymmetric vacuum we must require the 
vanishing of the SUSY variation of the gaugino.  Setting $X^{9}, X^{4}, X^{5}=0$ 
and 
taking into account the equations-of-motion for $D$, this condition reduces to 
\begin{equation}
\frac{dX^{a}}{ds}+\varepsilon^{abc}[X^{b},X^{c}]=-i\delta(s)\sigma^{a}_{ij}h_{i}
^{\ast}\otimes h_{j}~~, ~~a,b,c=1,2,3
\end{equation}
where $\sigma^{a}$ are the Pauli matrices.
Equation (8) is easily seen to reproduce Nahm's bc (Nahm's equations are just: 
left-hand-side of (8)=0) in the case of discontinuous but finite $X^{a}(s)$. 
The discontinuities in the mathematical literature are given in terms of a
rank-one $N\times N$ complex matrix which is parametrized by two complex 
$N$-vectors, the $u_{0}, u_{1}$ of [4]. These are essentially the $h_{1m}, 
h_{2m}$ 
of Eq.(8).
 
\section{Conclusion and Discussion}
The discontinuity conditions in Nahm's data were derived from D-brane 
considerations.

It should be noted that the $D^{2}$ term in the Langrangian apparently gives 
rise (on-shell) to an infinite-energy contribution. This is due to the singular 
nature of 
the configuration depicted in fig.1. However this picture is expected to get 
smoothed-out by the 1-3 sector strings in a similar to [10] manner. Another way 
of seeing this would be to ``lift'' the system to $M$ theory [11]. 

In Ref.[5] it was suggested that stringy contributions will give rise to a 
$\delta$ function  regularization of the form
\begin{displaymath}
\delta(s)\rightarrow \frac{1}{l_{s}}e^{-\frac{s^2}{l_{s}^{2}}}
\end{displaymath}
It would be interesting to see how can such stringy effects be taken into 
account in a systematic expansion in the string scale $l_{s}$. This problem will 
hopefully be investigated in some future work. 
\vspace{0.5cm}

{\bf Note}: While this paper was in the final stage of preparation an 
overlapping  publication [7] appeared.
\vspace{0.5cm}

\begin{center}
{\bf Acknowledgements}\\
\end{center}

I would like to thank Greg Moore for suggesting this problem to me, for numerous 
discussions and for a critical reading of the manuscript. I would also like to 
thank Ruben Minasian for discussions.

This work was supported by DOE grant DE-FG02-92ER40704.
\vspace{1cm}
\clearpage
{\bf References}

\begin{enumerate}

\item W. Nahm, {\it The Construction of All Self-Dual Multimonopoles by the ADHM 
Method}  in ``Monopoles in: Quantum Field Theory'', eds. Craigie et al. (World 
Scientific, Singapore, 1982); {\it A Simple Formalism for the BPS Monopole}, 
Phys. 
Lett. {\bf B90} (1980) 413; E. Corrigan, P. Goddard, {\it Construction of 
Instanton and Monopole Solutions and Reciprocity}, Ann. Phys. {\bf 154} (1984) 
253. 
\item J. Polchinski, {\it TASI Lectures on D-branes}, hep-th/9611050.
\item D.-E. Diaconescu, {\it D-branes, Monopoles and Nahm Equations}, hep-th/ 
9608163, Nucl. Phys. {\bf B503} (1997) 220.
\item J. Hurtubise, M.K. Murray, {\it On the Construction of Monopoles for the 
Classical Groups}, Commun. Math. Phys. {\bf 122} (1989) 35. 
\item A. Gerasimov, G. Moore, S. Shatashvilli, unpublished notes.
\item A. Giveon, D. Kutasov, {\it Brane Dynamics and Gauge Theory}, 
hep-th/9802067.
\item A. Kapustin, S. Sethi, {\it The Higgs Branch of Impurity Theories}, 
hep-th/9804027.
\item M.R. Douglas, {\it Gauge fields and D-branes}, hep-th/9604198.
\item P. West, {\it Introduction to Supersymmetry and Supergravity}, 2nd edition 
(World Scientific, Singapore,1990).
\item C. Callan, J. Maldacena, {\it Brane Dynamics from the Born-Infeld Action}, 
hep-th/9708147, Nucl. Phys. {\bf B513} (1998) 198.
\item E. Witten, {\it Solutions of Four Dimensional Field Theories via M 
Theory}, 
hep-th/9703166, Nucl. Phys. {\bf B500} (1997) 3.
\end{enumerate}
\clearpage

\setlength{\unitlength}{1cm}
\begin{picture}(4,4)(-5.8,-1)
\put(3,0){\vector(0,1){2}}
\put(3,0){\vector(1,0){2}}
\put(-3,-4){\line(0,1){6}}
\put(-6,-2){\line(1,0){6}}
\put(-6,0){\line(1,0){6}}
\put(-6,0.5){\line(1,0){6}}
\put(-5,-0.3){\circle*{0.1}}
\put(-5,-0.6){\circle*{0.1}}
\put(-5,-0.9){\circle*{0.1}}
\put(6,1.){\circle*{0.1}}
\put(6,1.){\circle{0.3}}
\put(-6.4,0.5){\mbox{1}}
\put(-6.4,0){\mbox{2}}
\put(-6.4,-2){\mbox{N}}
\put(0,-2.5){\mbox{D1}}
\put(-2.7,1.7){\mbox{D3}}
\put(6.3,1.){\mbox{$x^{4-8}$}}
\put(5,0.2){\mbox{$x^{9}$}}
\put(3.2,2){\mbox{$x^{1-3}$}}
\put(-6.0,-4.9){\mbox{{\bf figure 1.}  A D3-brane extended in $x^{0,1,2,3}$ intersecting a system of $N$}}
\put(-4.2,-5.3){\mbox{parallel D1-branes. The intersection is at the point $x^{9}=0$}}
\put(-4.2,-5.7){\mbox{on the D1-branes' world-volume.}}

\end {picture}
\end{document}